\documentclass[pre, twocolumn, longbibliography, amsfonts, amssymb, amsmath, nobalancelastpage]{revtex4-1}

\usepackage{mathbbol,amsthm}
\usepackage{graphicx}

\DeclareMathOperator*{\argmax}{argmax}

\newtheorem{thm}{Theorem}

\newcommand{\Rr}{\mathbb{R}}

\newcommand{\R}{\mathcal{R}}

\renewcommand{\L}{\mathcal{L}}
\renewcommand{\S}{\mathcal{S}}

\newcommand{\G}{\mathcal{G}}
\newcommand{\Hg}{\mathcal{H}}
\newcommand{\V}{\mathcal{V}}
\newcommand{\E}{\mathcal{E}}

\begin{document}

\title{Alignment and integration of complex networks by
  hypergraph-based spectral clustering}

\author{Tom Michoel}

\email{tom.michoel@roslin.ed.ac.uk}

\affiliation{Freiburg Institute for Advanced Studies (FRIAS),
  University of Freiburg, Albertstrasse 19, D-79104 Freiburg, Germany}

\affiliation{The Roslin Institute, The University of Edinburgh, Easter
  Bush, Midlothian, EH25 9RG, Scotland, UK}

\author{Bruno Nachtergaele}

\email{bxn@math.ucdavis.edu}

\affiliation{Department of Mathematics, University of California
  Davis, One Shields Avenue, Davis, CA 95616-8366, USA}

\begin{abstract} 
  Complex networks possess a rich, multi-scale structure reflecting
  the dynamical and functional organization of the systems they
  model. Often there is a need to analyze multiple networks
  simultaneously, to model a system by more than one type of
  interaction or to go beyond simple pairwise interactions, but
  currently there is a lack of theoretical and computational methods
  to address these problems. Here we introduce a framework for
  clustering and community detection in such systems using hypergraph
  representations. Our main result is a generalization of the
  Perron-Frobenius theorem from which we derive spectral clustering
  algorithms for directed and undirected hypergraphs. We illustrate
  our approach with applications for local and global alignment of
  protein-protein interaction networks between multiple species, for
  tripartite community detection in folksonomies, and for detecting
  clusters of overlapping regulatory pathways in directed networks.
\end{abstract}

\maketitle

\section{Introduction}

Complex networks in nature and society represent interactions between
entities in inhomogeneous systems and understanding their structure
and function has been the focus of much research. At the macroscopic
scale, complex networks are characterized by, among others, a degree
distribution, characteristic path length and clustering coefficient
which are markedly different from those of regular lattices or
uniformly distributed Erd\H{o}s-R\'enyi random graphs
\cite{newman2003,albert2002}, while at the microscopic scale, they
contain network motifs, small subgraphs occurring significantly more
often than expected by chance \cite{milo2002}.  The intermediate level
usually exhibits the presence of communities or modules, sets of nodes
with a significantly higher than expected density of links between
them, typical examples being friendship circles in social networks,
websites devoted to similar topics in the World Wide Web or protein
complexes in protein interaction networks \cite{newman2006b,
  fortunato2009, leskovec2009community,porter2009}.

However, the limitations of modeling a complex system by a network
with a single type of pairwise interaction are becoming more and more
clear. Folksonomies, online social communities where users apply tags
to annotate resources such as images or scientific articles, have a
tripartite structure with three types of interactions
\cite{ghoshal2009random, zlatić2009hypergraph}.  In biology, cellular
systems are characterized by different types of networks which
represent different physical interaction mechanisms operating on
different time-scales, intertwined with each other through extensive
feedforward and feedback loops \cite{alon2007b,zhu2007}. To understand
how evolutionary dynamics shapes molecular interaction networks, we
need to compare them between multiple species with non-trivial
many-to-many relations between their respective node sets
\cite{sharan2006}. In order to move beyond simple networks of pairwise
interactions to model these and other systems, one suggestion has been
to use hypergraphs, where edges are arbitrarily sized subsets of
nodes.  Although a number of studies have generalized various concepts
from graph theory to hypergraphs \cite{zhou2007,
  vazquez2008population, klamt2009, vazquez2009finding,
  ghoshal2009random, zlatić2009hypergraph}, a rigorous mathematical
foundation and general-purpose algorithm for clustering and community
detection in hypergraphs is still lacking.

Here we present a framework for spectral clustering in hypergraphs
which is mathematically sound and algorithmically efficient. It is
based on a generalization of the Perron-Frobenius theorem, which
allows to define and compute a dominant eigenvector for hypergraphs
and use its values for optimally partitioning the hypergraph's vertex
set, similar to the operation of standard spectral clustering
algorithms in ordinary graphs \cite{newman2006}.  We demonstrate the
validity of our approach through practical applications in the
analysis of real-world networks. In particular we address the
following problems. First, if two networks are defined on separate
node sets with a many-to-many mapping between them (for instance
protein-protein interaction networks in different species), it is a
natural question to find matching communities in the two
networks. This is the so-called \textit{network alignment} problem
\cite{sharan2006}. We show that this problem can be solved by finding
clusters in a hypergraph where each hyperedge consists of two matching
edges, one from each network (Section
\ref{sec:local-glob-alignm}). Second, if multiple networks are defined
on the same node set (\textit{i.e.}, together they form an
edge-colored graph), there often exist functionally meaningful,
higher-order relations between the different edge types (for instance
tripartite relations in folksonomies \cite{ghoshal2009random,
  zlatić2009hypergraph} or network motifs in biological networks
\cite{alon2007b,zhu2007}). Finding communities or modules which
respect to these higher-order relations is what we call the
\textit{network integration} problem. Here we show that any
higher-order edge relation between different networks defines a
subgraph pattern in the corresponding edge-colored graph and that all
instances of this pattern form a hypergraph. Hypergraph-based
clustering can then be applied to identify modules in such
edge-colored graphs (Section \ref{sec:tripartite} and
\ref{sec:path-clust-regul}).

\section{Graphs and hypergraphs}
\label{sec:graphs-hypergr-tens}

A graph $\G$ is defined as a pair $(\V,\E)$ of vertices $\V$ and edges
(pairs of vertices) $\E$, which may be directed or not. In a weighted
graph, a number is assigned to each edge which may represent,
\textit{e.g.}, the cost, length or reliability of an edge. A
hypergraph is a generalization of a graph where an edge, called
hyperedge in this case, can connect any number of vertices,
\textit{i.e.}, $\E$ is a set of arbitrarily sized subsets of $\V$. A
particular class of hypergraphs are so-called $k$-uniform hypergraphs
where each hyperedge has the same cardinality $k$.  Algebraically, a
graph can be represented by an adjacency matrix $A$ of dimension
$N\times N$, with $N$ the number of vertices, such that $A_{ij}=1$ if
$\{i,j\}\in \E$ and $0$ otherwise. For undirected graphs, $A$ is a
symmetric matrix and for weighted graphs, $A_{ij}$ is defined to be
the weight of the edge $\{i,j\}$. For $k$-uniform hypergraphs, the
notion of adjacency matrix can be generalized to an adjacency
multi-array or tensor $T$, with $T_{i_1\dots i_k}=1$ if $\{i_1,\dots,
i_k\}\in \E$ and $0$ otherwise.  For a general hypergraph, we define a
function $w$ on the set of subsets of $\V$ such that $w(E)=1$ for
$E\in\E$ and $0$ otherwise. In general, we allow weighted hypergraphs
where $w$ can be any non-negative function.

A path between two vertices $i$ and $j$ in a hypergraph is defined as
a sequence of vertices $i=i_1,i_2,\dots, i_{k+1}=j$ and edges
$E_1,\dots, E_k$ such that for all $m$, $\{i_m,i_{m+1}\}\subset
E_m$. A hypergraph is called \emph{connected} if there exists a path
between any pair of vertices. A stronger constraint on the structure
of a hypergraph is that of \emph{irreducibility}.  A hypergraph is
said to be reducible if there exists a proper vertex subset $I\subset
\V$ such that for any $i\in I$ and $j_1,\dots, j_m \not\in I$,
$w\bigl(\{i,j_1,\dots,j_m\}\bigr)=0$, and irreducible if it is not
reducible. For ordinary graphs, connectedness and irreducibility are
equivalent, but for hypergraphs this is not the case. An irreducible
hypergraph is clearly connected, but the opposite is not always
true. Indeed, if there exists a subset of vertices $I$ such that paths
crossing from $i\in I$ to $j\not\in I$ can always be chosen to do so
through an edge of the form $\{i_1,\dots,i_k,j_1,\dots,j_m\}$, with
$k\geq 2$, $i_1,\dots,i_k\in I$ and $j_1,\dots,j_m\not\in I$, then we
can set $w(\{i,j_1,\dots,j_m\})=0$ for all $i\in I$ and
$j_1,\dots,j_m\not\in I$, thereby making the hypergraph reducible,
without breaking its connectivity.

Directed hypergraphs can be defined in many ways. For instance for
$k$-uniform hypergraphs, we can impose any form of permutation
symmetry, or lack thereof, between some or all of the $k$ dimensions
in each edge. In this paper, we will only consider the case where each
edge $E$ can be written as a pair $(S,T)$, where $S\subset\V$ is
called the `source' vertex set and $T\subset\V$ the `target' vertex
set, with weight function $w(S,T)$. Underlying a directed hypergraph,
there is always an undirected hypergraph with edges $E=S\cup T$ for
every directed edge $(S,T)$. As is the case for ordinary directed
graphs, a stronger notion of connectivity is usually needed than
simple connectivity of this undirected hypergraph. We defer the
somewhat technical definition of strong connectivity of directed
hypergraphs to Appendix \ref{sec:strong-conn-direct}.

\section{Dominant eigenvectors and spectral graph clustering}
\label{sec:domin-eigenv-spectr}

Although countless measures have been designed to define clusters in a
graph \cite{fortunato2009, leskovec2009community,porter2009}, perhaps the simplest definition
is that a cluster is a subset of vertices with a high number of edges
between them, relative to its size. Mathematically, for a graph with
adjacency matrix $A$, the edge-to-node ratio of a subset $X\subset \V$
can be written as
\begin{align*}
  \S(X) =  \frac{\sum_{i,j\in X} A_{ij}}{|X|},
\end{align*}
where $|X|$ denotes the number of elements in $X$. The number of
subsets of a set with $N$ elements grows exponentially in $N$ and
hence finding the subset with maximal edge-to-node ratio by
exhaustive enumeration is computationally infeasible for large
graphs. However, if we denote by $u_X$ the unit vector in $\Rr^{N}$
which has $u_{X,i}=|X|^{-1/2}$ for $i\in X$ and $0$ otherwise, we can
write $\S$ as a scalar product and obtain the simple upper bound:
\begin{equation}\label{eq:12}
  \S(X) =\langle u_X, A u_X\rangle \leq \max_{x\in\Rr^N, x\neq 0} \frac{\langle
    x,Ax\rangle}{\|x\|^2} = \lambda_{\max},
\end{equation}
where $\langle x,y\rangle=\sum_i x_i y_i$ is the standard inner
product on $\Rr^N$, $\|x\|=\sqrt{\langle x,x\rangle}$ is the length of
$x$, and $\lambda_{\max}$ is the largest eigenvalue of $A$. By the
Perron-Frobenius theorem \cite{horn1985}, if the graph is irreducible,
the dominant eigenvector $x$, which satisfies $\lambda_{\max}\, x =
Ax$, is unique, strictly positive ($x_{i}>0$ for all $i$), and solves
the variational problem in the right-hand side of eq. \eqref{eq:12}.

Hence, to find an approximate maximizer $X$ of $\S$, we can take the
set $X$ for which $u_X$ is as close as possible to the dominant
eigenvector $x$, similar to what is done in other spectral clustering
algorithms based on the Laplacian or modularity matrices
\cite{newman2006}, \textit{i.e.}, define
\begin{align*}
  \tilde X &= \argmax_{X\subset \V} \langle u_X, x\rangle =
  \argmax_{X\subset \V} \frac1{|X|^{1/2}} \sum_{i\in X} x_{i}.
\end{align*}
Since $x>0$, $\tilde X$ is of the form $ X_c = \{ i\colon x_i>c\}$ for
some threshold value $c$. Instead of $\tilde X$, we therefore choose
the solution of the restricted variational
problem
\begin{equation}
  \label{eq:8}
  X_{\max} = \argmax_{c>0} \S(X_c).
\end{equation}
as an approximate maximizer.  Solving eq. \eqref{eq:8} is linear in
the number of vertices, since we only need to consider the values $c$
equal to the entries of $x$. Moreover, $\S(X_{\max})\geq \S(\tilde
X)$, and hence $X_{\max}$ is a better approximation to the true
maximizer of $\S$ than $\tilde X$.

Thus we obtain a numerically highly efficient spectral graph
clustering algorithm:
\begin{enumerate}
\item Calculate the dominant eigenvector $x$ using for instance a
  power method \cite{golub1996}.
\item Find the cluster $X_{\max}$ which solves the restricted
  variational problem in eq. \eqref{eq:8}.
\item Store $X_{\max}$, remove all edges between nodes in $X_{\max}$
  from the edge set $\E$, and repeat the procedure until no more edges
  remain.
\end{enumerate}
This result of this algorithm is a partition of the edges of the input
graph. Edge clustering algorithms have recently gained popularity as
they allow for overlapping communities where nodes may belong to more
than one community \cite{evans2009line,ahn2010link}.

This procedure generalizes immediately to directed or bipartite
graphs. In this case a cluster consists of a `source' set $X$ and
`target' set $Y$ with edge-to-node ratio
\begin{align*}
  \S(X,Y) = \frac{\sum_{i\in X, j\in Y}A_{ij}}{\sqrt{|X|\cdot |Y|}}.
\end{align*}
The dominant eigenvector is replaced by the dominant left and right
singular vectors $x$ and $y$ corresponding to the largest singular
value of $A$, which are again unique and strictly positive
\cite{horn1985}. $X_{\max}$ and $Y_{\max}$ are found by maximizing
$\S(X,Y)$ over sets obtained by thresholding on the entries of $x$ and
$y$.

\section{Perron-Frobenius theorem for hypergraphs}
\label{sec:perr-frob-hypergr}

Our aim is to generalize the previous graph spectral clustering
algorithm to arbitrary hypergraphs. For this purpose we first need a
generalization of the Perron-Frobenius theorem.  Let $\Hg=(\V,\E)$ be
an undirected hypergraph on $N$ vertices. Define for $x\in\Rr^N$ and
$p\geq1$
\begin{equation}\label{eq:6}
  \R_p(x) = \sum_{E\in\E} w(E) \prod_{i\in E} \Bigl(\frac{|x_i| }
  {\|x\|_p}\Bigr)^{\frac1{|E|}},
\end{equation}
where $w(E)$ is the non-negative weight of edge $E$ and $\|x\|_p =
(\sum_i |x_i|^p)^{1/p}$ is the $p$-norm of $x$.  We have the following
key result:
\begin{thm}\label{thm:perron-frobenius} 
  $\R_p$ attains its maximum on the set of unit vectors
  $\mathbb{S}^N_p = \{u\in\Rr^N\colon \|u\|_p=1\}$. If $\Hg$ is
  connected, there is a unique maximizer $x\in\mathbb{S}^N_p$ which is
  strictly positive and satisfies the Euler-Lagrange equations
  \begin{equation}\label{eq:5}
    \lambda_p\, x_i^{p} = \sum_{\{E\in\E\colon i\in E\}}
    \frac{w(E)}{|E|} \Bigl(\prod_{j\in E} x_{j}\Bigr) ^{\frac1{|E|}}, 
  \end{equation}
  subject to the constraint $\|x\|_p=1$ and with $\lambda_p=
  \R_p(x)$. By analogy with the matrix case, we call $x$ the dominant
  eigenvector of $\Hg$.
\end{thm}
For clarity, we first prove this theorem in the simpler case when
$\Hg$ is irreducible. The proof of the general case is given in
Appendix \ref{sec:general-proof-perron}.
\begin{proof}
  Existence of a maximizer on $\mathbb{S}^N_p$ follows from
  Weierstrass's theorem \cite{horn1985}.  Clearly, since
  $\R_p(x)=\R_p(|x|)$, we can always choose a maximizer $x$ to have
  non-negative entries. Hence we can find $x$ as a stationary point of
  the Lagrangian
  \begin{align*}
    \L(x) = \sum_{E\in\E} w(E) \Bigl(\prod_{i\in E}
    |x_i|\Bigr)^{\frac1{|E|}} - \frac{\lambda}{p} \bigl( \|x\|_p^p -1
    \bigr),
  \end{align*}
  giving rise (for non-negative $x$) to the Euler-Lagrange equations
  \begin{equation}\label{eq:10}
    \lambda x_i^{p-1} = \sum_{\{E\in\E\colon i\in E\}} \frac{w(E)}{|E|} \Bigl(\prod_{j\in E,
      j\neq i} x_{j}\Bigr) ^{\frac1{|E|}} x_i^{\frac1{|E|}-1}.
  \end{equation}
  Let $I=\{i\in\V\colon x_i=0\}$ and $i\in I$. Assume there exists an
  edge $E=\{i,j_1,\dots,j_m\}$ with $j_1,\dots,j_m\not\in I$. Then the
  left-hand side of eq. \eqref{eq:10} is $0$ while the right-hand side
  is $\infty$. Hence such an edge cannot exist, but this contradicts
  the assumption of irreducibility of $\Hg$. It follows that
  $I=\emptyset$ or $x>0$. Multiplying both sides of eq. \eqref{eq:10}
  by $x_i$ we obtain eq. \eqref{eq:5}. Summing both sides in
  eq. \eqref{eq:5} over $i$ gives $\lambda_p= \R_p(x) =
  \max_{x'}\R_p(x')$.
  
  Next assume $y>0$ is another maximizer of $\R_p$. Denote
  $c=\min_i(x_{i}/y_{i})$, $u=cy$, and $z=x-u\geq 0$. Since
  $\|x\|_p=\|y\|_p=1$, we have $c<1$ and $c^{p}\leq c$ for $p\geq
  1$. Denote $I=\{i\in\V\colon z_i=0\}$. For any $i\in I$, by the
  Euler-Lagrange equations,
  \begin{multline*}
    0 = \lambda_p \bigl(x_{i}^{p} -c^{p} y_{i}^{p} \bigr) \\
    \geq \sum_{\{E\in\E\colon i\in E\}} w(E) \Bigl[\Bigl(\prod_{j\in
      E} x_{j}\Bigr)^{\frac1{|E|}} - \Bigl(\prod_{j\in E}
    u_{j}\Bigr)^{\frac1{|E|}}\Bigr].
  \end{multline*}
  Since each term in the last sum is non-negative, they must all be
  zero. Hence for any $j_1,\dots,j_k\not\in I$, if
  $\{i,j_1,\dots,j_k\}\in\E$ then
  \begin{align}
    0 &= \prod_{m=1}^k x_{j_m}  - \prod_{m=1}^k u_{j_m} \nonumber\\
    &= \sum_{m=1}^k \bigl(\prod_{n=1}^{m-1} u_{j_n}\bigr)
    (x_{j_m}-u_{j_m}) \bigl(\prod_{n=m+1}^{k} x_{j_n}\bigr).\label{eq:2}
  \end{align}
  Again each term in this sum is non-negative and must therefore be
  zero, but this contradicts $j_1,\dots,j_k \not\in I$. Hence edges
  with $i\in I$ and $j_1,\dots,j_k \not\in I$ do not exist, but this
  contradicts the assumption of irreducibility. Since
  $I\not=\emptyset$, we must have $I=\mathcal{V}$ or $x=y$.
\end{proof}

Next consider directed hypergraphs with hyperedges $E=(S,T)$,
$S,T\subset\V$ as defined before. Then define $\R_{p,q}(x,y)$ for
$x,y\in\Rr^N$ and $p,q\geq 1$
\begin{multline}\label{eq:7}
  \R_{p,q}(x,y) =\\ \sum_{(S,T)\in\E} w(S,T) \prod_{i\in S} \Bigl(\frac{|x_i| }
  {\|x\|_p}\Bigr)^{\frac1{2|S|}}\prod_{j\in T}\Bigl(\frac{|y_j| }
  {\|y\|_q}\Bigr)^{\frac1{2|T|}}.
\end{multline}
By identical arguments as for undirected hypergraphs, it can be shown
that for a strongly connected directed hypergraph, there exists a
unique pair $x\in\mathbb{S}^N_p$ and $y\in\mathbb{S}^N_q$ such that
$\R_{p,q}(x,y)\geq \R_{p,q}(x',y')$ for all $x',y'\in\Rr^N$. These
maximizers are strictly positive and satisfy the Euler-Lagrange
equations
\begin{align}
  \lambda_{p,q} x_i^p &= \sum_{\{(S,T)\in\E\colon i\in S\}}
  \frac{w(S,T)}{2|S|} \Bigl(
  \prod_{i'\in S} x_{i'} \Bigr)^{\frac1{2|S|}}\Bigl(\prod_{j\in  T}y_j\Bigr)^{\frac1{2|T|}}\label{eq:9}\\
  \lambda_{p,q} y_j^q &= \sum_{\{(S,T)\in\E\colon j\in T\}}
  \frac{w(S,T)}{2|T|} \Bigl( \prod_{i\in S} x_i
  \Bigr)^{\frac1{2|S|}}\Bigl(\prod_{j'\in
    T}y_{j'}\Bigr)^{\frac1{2|T|}},\label{eq:11}
\end{align}
subject to the constraints $\|x\|_p=\|y\|_q=1$ and with
$\lambda_{p,q}=\R_{p,q}(x,y)$. Details are given in Appendix
\ref{sec:general-proof-perron}.

\section{Spectral clustering and biclustering in hypergraphs}
\label{sec:spectr-clust-bicl}

Having a generalization of the Perron-Frobenius theorem, it is
straightforward to also generalize the spectral clustering
method. Define for $X\subset \V$,
\begin{equation}\label{eq:1}
  \S_p(X) = \frac{\sum_{E\subset X} w(E)}{|X|^{\frac1p}} =
  \R_p(u_X)\leq \R_p(x),
\end{equation}
with $x$ the dominant eigenvector and $u_X\in\mathbb{S}_p^N$ now
defined by $u_{X,i}=|X|^{-1/p}$ for $i\in X$ and $0$ otherwise. The
parameter $p$ balances cluster size versus edge density. For $p=1$,
$\S_p$ is the ratio of edges to nodes in $X$. Taking $p>1$ diminishes
the influence of the denominator and progressively favors to have a
high number of edges rather than a high number of edges per node in
high-scoring clusters (further details in Section
\ref{sec:algorithm-validation}). The spectral clustering algorithm
becomes:
\begin{enumerate}
\item Calculate the maximizer $x$ of $\R_p$.
\item Find the cluster $X_{\max}$ which solves the restricted
  variational problem
  \begin{align*}
    X_{\max} = \argmax_{c>0} \S_p(X_c)
  \end{align*}
  with $X_c=\{i\in\V\colon x_i>c\}$.
\item Store $X_{\max}$, remove all hyperedges between nodes in
  $X_{\max}$ from the edge set $\E$, and repeat the procedure until no
  more hyperedges remain.
\end{enumerate}
The maximizer can be calculated using a generalization of the power
method for matrices \cite{golub1996} or tensors
\cite{delathauwer2000}: starting with an initial vector $x^{(0)}$ and
defining $\lambda^{(0)}_p=\|x^{(0)}\|_p=1$, we
compute $x^{(n+1)}$ from $x^{(n)}$ using the Euler-Lagrange equations
\eqref{eq:5} in the following steps:
\begin{align}
  x^{(n+1)}_i &\leftarrow \biggl[ \sum_{\{E\in\E\colon i\in E\}} \frac{w(E)}{|E|}
  \Bigl(\prod_{j\in E} x_{j}^{(n)}\Bigr)
  ^{\frac1{|E|}}\biggr]^{\frac1{p}} \label{eq:13}\\
  \lambda^{(n+1)}_p &= \|x^{(n+1)}\|_p \label{eq:14}\\
  x^{(n+1)}_i &\leftarrow \frac{x^{(n+1)}_i}{\lambda^{(n+1)}_p}, \label{eq:15}
\end{align}
iterated until the components of $x^{(n)}$ become stationary or,
equivalently, $\lambda^{(n)}_p$ has converged to the dominant
eigenvalue, \textit{i.e.},
\begin{equation}\label{eq:16}
  \left|1-\frac{\lambda^{(n+1)}_p}{\lambda^{(n)}_p}\right|<\epsilon,
\end{equation}
where $\epsilon$ is a predefined numerical tolerance threshold. Due to
the uniqueness of $x$, the choice of starting vector is not
important. By taking a non-negative one, such as the uniform vector
$x^{(0)}=[1,1,\dots,1]^T/N^{1/p}$, we ensure that the powers of
$1/|E|$ occurring in the Euler-Lagrange equations are always defined
unambiguously. Many of the hypergraphs occurring in real-world
applications are not connected. In such cases it is important to
ensure that $x^{(0)}$ has support only on a single connected component
to obtain the unique maximizer for that component.

Although we typically view a cluster as a subset of vertices, it is
actually a subset of hyperedges (all hyperedges $E\subset X_{\max}$)
and thus can be considered as a sub-hypergraph as well.
Higher-scoring clusters can thus be obtained by recursively applying
the previous procedure to each of the clusters itself until no more
subdivision which improves the score is found.

For directed hypergraphs, we have a biclustering method. Define for
$X,Y\subset\V$ and $p,q\geq 1$
\begin{align*}
  \S_{p,q}(X,Y) = \frac{\sum_{S\subset X,T\subset Y}
    w(S,T)}{|X|^{\frac1{2p}} |Y|^{\frac1{2q}}}.
\end{align*}
Approximate maximizers $X_{\max}$ and $Y_{\max}$ are found by
solving the restricted variational principle
\begin{align*}
  (X_{\max},Y_{\max}) = \argmax_{(c_1,c_2)} \S_{p,q}(X_{c_1},Y_{c_2}),
\end{align*}
with $X_{c_1}=\{i\in\V\colon x_i>c_1\}$ and $Y_{c_2}=\{i\in\V\colon y_i>c_2\}$,
where $x$ and $y$ are the unique solutions of the Euler-Lagrange
equations \eqref{eq:9}-\eqref{eq:11}, which can again be calculated
using a power algorithm.

\section{Relation to previous work}

The matrix algorithm for clustering in a simple graph has its roots in
a method for image pattern recognition \cite{inoue1999} and using the
singular value decomposition to detect densely linked sets in directed
networks goes back to the work of Kleinberg \cite{kleinberg1999}.  The
novelty here lies in the definition of a discrete cluster through
solving the restricted variational problem, instead of using an ad-hoc
cut-off on the eigenvector entries.  For $k$-uniform hypergraphs, we
can define rescaled variables $y_i = x_i^{1/k}$ such that maximizing
$\R_p(x)$ becomes equivalent to maximizing
\begin{align*}
  \R'_{p'}(y) = \frac{\sum_{i_1,\dots,i_k} T_{i_1\dots i_k} y_{i_1}\dots
    y_{i_k}}{\|y\|_{p'}^k}
\end{align*}
with $p'=kp$. In this case, Theorem \ref{thm:perron-frobenius} reduces
to a multi-linear extension of the Perron-Frobenius theorem to
non-negative irreducible tensors of arbitrary dimension, which has
been the subject of several recent papers
\cite{lim2005,chang2008,friedland2011} (which all depend on the strong
irreducibility condition). The proof given in Theorem
\ref{thm:perron-frobenius} is considerably simpler, holds for general
connected hypergraphs and follows more closely the proof of the matrix
theorem \cite{horn1985}.  In the unscaled variational problem for
$\R'_{p'}$, the maximizer is unique for $p'\geq k$ and thus it is
unsuited for generalizing to arbitrary hypergraphs where the
uniqueness condition would become $p'\geq k_{\max}$, the maximum edge
size in the hypergraph. This explains why we introduced the geometric
average over the values $x_j$ in eq. \eqref{eq:6}.  To the best of our
knowledge, Theorem \ref{thm:perron-frobenius} is the first proof of a
Perron-Frobenius theorem for general hypergraphs.

For $k=3$, we have previously used a similar approach to find clusters
of $3$-node network motifs in integrated interaction networks
\cite{michoel2011,audenaert2011}. In this case an adjacency tensor $T_{rst}$ is
defined to be $1$ if an instance of a $3$-node query motif or graph
pattern exists between vertices $(r,s,t)$ and $0$ otherwise. More
generally, we can define for any $k$-node query pattern a $k$-uniform
hypergraph consisting of all instances of the query pattern in a given
graph $\G$. Our algorithm will identify clusters of vertices in $\G$
with a high number of pattern instances between them, which often have
a functional meaning in biological networks
\cite{michoel2011,zhang2005}.

Another example for $k=3$ concerns the analysis and clustering of
multiply linked data \cite{dunlavy2006,li2011} or multislice networks
\cite{mucha2010community}. Here we are given a set of $M$ directed or
undirected graphs and define a hypergraph adjacency tensor as $T_{ijm}
= A_{ij}^{(m)}$, where $A^{(m)}$ denotes the adjacency matrix of the
$m^{\mathrm{th}}$ graph. Clustering in this case identifies vertex
sets which are densely connected in multiple, but not necessarily all,
graphs.

\section{Algorithm validation}
\label{sec:algorithm-validation}

\subsection{Random geometric graphs}
\label{sec:rand-geom-graphs}

The dominant eigenvector of a graph's adjacency matrix is often
considered as a centrality measure (`eigenvector centrality'
\cite{newman2003}) and is, in essence, equal to a simplified PageRank
\cite{brin1998} for ranking global vertex importance. It may thus come
as a surprise to see it playing a role in identifying localized
clusters (however, see the references in the previous section). In
order to demonstrate the validity of our approach and illustrate the
statements in Section \ref{sec:spectr-clust-bicl}, we applied it to
randomly generated geometric graphs of various sizes (see Appendix
\ref{sec:geom-rand-graphs} for details).

\begin{figure*}
  \centering
  \includegraphics[width=\linewidth]{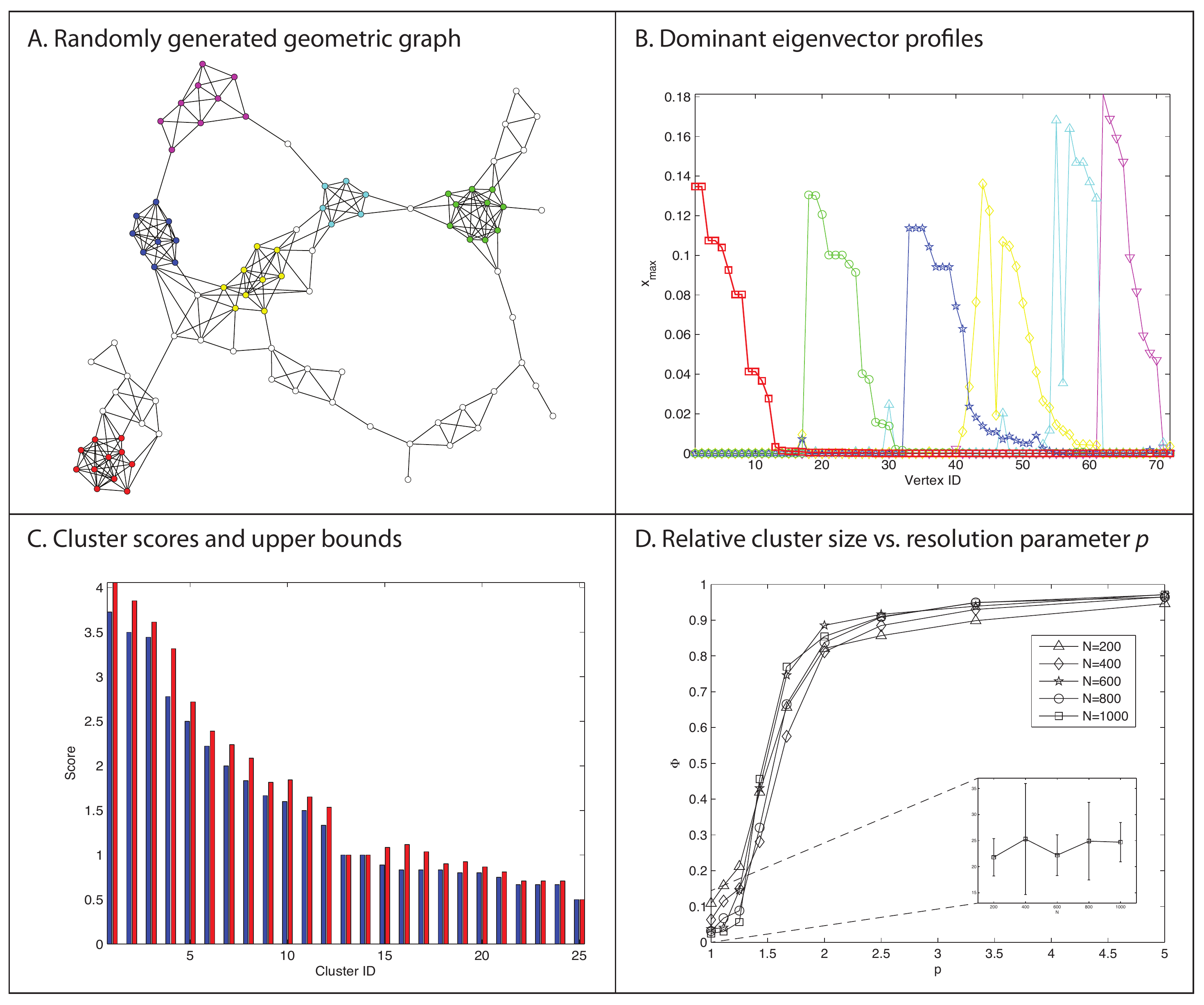}
  \caption{(Color online) \textbf{A.} Example of a randomly generated
    geometric graph with 100 vertices and radius $r^2=0.02$, showing
    the largest connected component with the six highest-scoring edge
    clusters indicated by filled nodes. \textbf{B.} Dominant
    eigenvector profiles for the six highest-scoring edge
    clusters. \textbf{C.} Edge-to-node ratio scores (left blue bars)
    and theoretical upper bound (right red bars) for all 25 edge
    clusters. \textbf{D.}  Cluster size as the fraction $\Phi$ of
    total number of network nodes for the highest-scoring
    triangle-based cluster in random geometric graphs with $N=$ 200,
    400, 600, 800 and 1000 nodes and constant edge density ($\rho=4$)
    as a function of $p$. Each data point is an average over 10 random
    networks. The insert shows the absolute mean cluster size and
    standard deviation over 10 random networks as a function of $N$
    for $p=1$.}
  \label{fig:geom100}
\end{figure*}

For visualization purposes, we generated as a toy example a random
geometric graph with $100$ vertices and radius $r^2=0.02$
(Fig. \ref{fig:geom100}A). The graph is evidently modular and the six
highest-scoring edge clusters identified by our algorithm (with $p=1$)
are indicated in color. The profiles of the corresponding dominant
eigenvectors are clearly localized on a subset of nodes
(Fig. \ref{fig:geom100}B), illustrating that in a modular network, the
dominant eigenvector indeed indicates the location of a single
cluster. Furthermore, comparing the edge-to-node ratio for each of the
discovered edge clusters with the theoretical upper bound in
eq. \eqref{eq:12} shows that the solution of the restricted
variational problem (eq. \eqref{eq:8}) must be close to the true
maximum (Fig. \ref{fig:geom100}C).

For a more systematic analysis we performed triangle-based clustering
on sequences of geometric graphs with constant expected edge density
and varying size. Triangle-based clustering searches for overlapping
sets of triangles in an ordinary graph and corresponds to the simplest
form of $k$-clique clustering \cite{palla2005}. Here we considered
each instance of a triangle in the input graph as a hyperedge in a
3-uniform hypergraph to which we applied our spectral clustering
algorithm.  The parameter $p$ can be used to identify clusters at
different levels of resolution. Independent of network size, there is
a low-$p$ phase where the fraction of nodes in a cluster is small
compared to total network size, and a high-$p$ phase where a cluster
consists of a macroscopic network portion (Fig.~\ref{fig:geom100}D).
Interestingly, at $p=1$ cluster size does not depend on network size
(Fig. ~\ref{fig:geom100}D, insert). Hence clustering based on
(hyper)edge-to-node ratio scores (eq. \eqref{eq:1}) does not suffer
from a resolution limit problem where cluster size grows with network
size irrespective of the presence of `natural' clusters at smaller
scales \cite{fortunato2007,good2010}.  As in the previous example, the
cluster scores are always close to their theoretical upper bounds,
demonstrating that the solution of the restricted variational problem
is close to the true optimum in all cases
\cite[Fig. S1]{hypergraphclust_pre_supp}.

\subsection{Edge-to-node scaling parameter}
\label{sec:edge-node-scaling}

\begin{figure*}
  \centering
  \includegraphics[width=\linewidth]{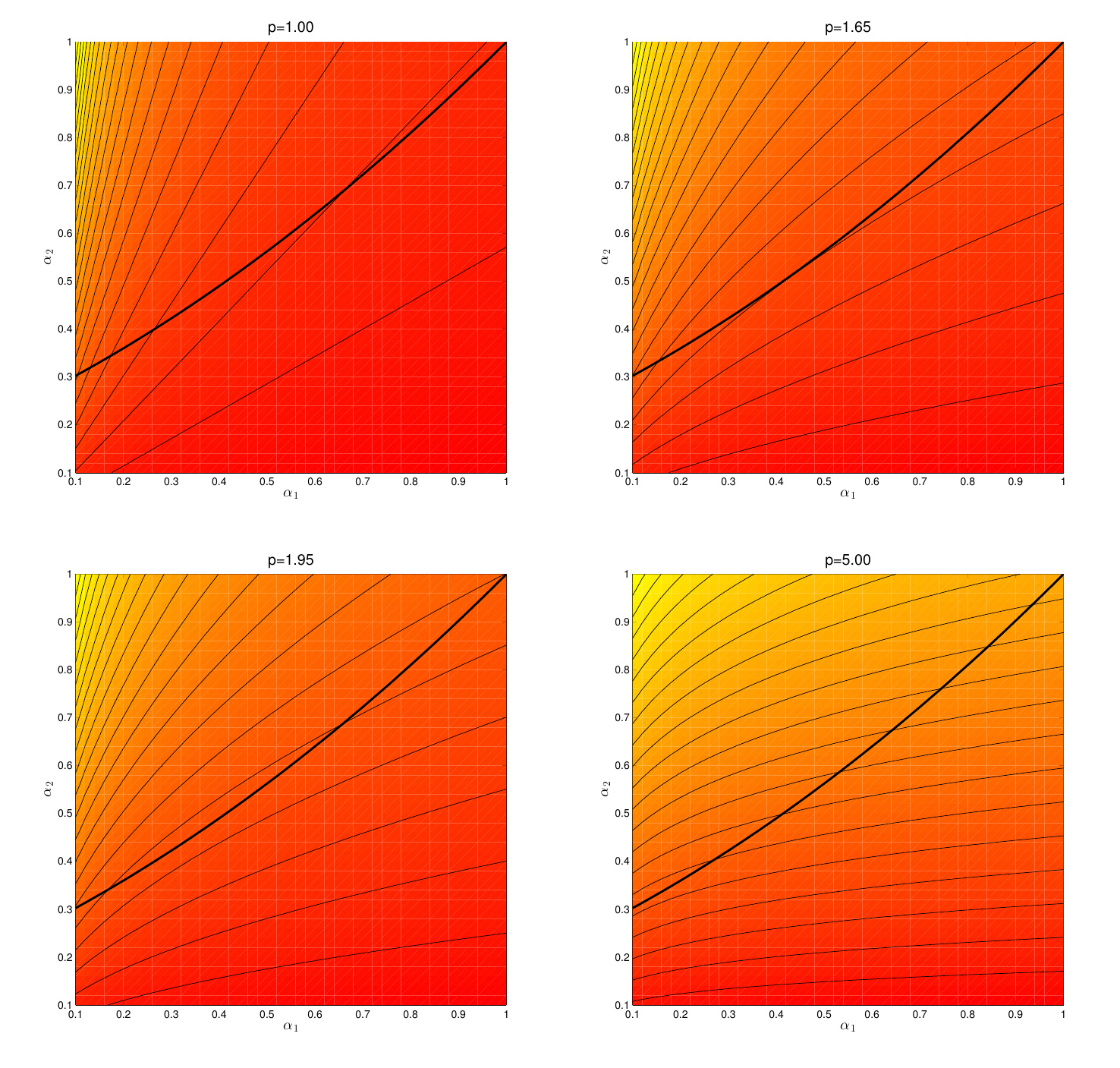}
  \caption{(Color online) Phase diagrams of $s_p(\alpha_1,\alpha_2)$
    for $p=1,1.65,1.95$ and $5$ (left to right, top to bottom). More
    yellow (lighter gray) indicates higher values of $s_p$; the thin
    lines are contours of constant $s_p$, while the thick line
    indicates a possible boundary of admissible states. Colors (gray
    scale levels) are relative to the minimum and maximum in each
    panel and not comparable between panels.}
  \label{fig:phase-diagrams}
\end{figure*}

The transition in Fig. ~\ref{fig:geom100}D as a function of the
edge-to-node scaling parameter $p$ is a general feature, independent
of the actual hypergraphs used, and can be easily understood as
follows. Assume we have a hypergraph $\Hg=(\V,\E)$ with $N=|\V|$ nodes
and $M=|\E|$ hyperedges. Then the relative score of any set
$X\subset\V$ with $n=|X|$ nodes and $m$ hyperedges compared to the
score of the total hypergraph is
\begin{align*}
  \frac{\S_p(X)}{\S_p(\V)} = \frac{m}{M}\left(\frac{N}{n}\right)^{\frac{1}{p}}
  = \frac{\alpha_2}{\alpha_1^{1/p}}\equiv s_p(\alpha_1,\alpha_2),
\end{align*}
with $\alpha_1$ and $\alpha_2$ de fractions of nodes and edges in
$X$. The phase diagram of $s_p$ as a function of these two variables
is independent of the actual hypergraph under consideration
(Fig. \ref{fig:phase-diagrams}). Naturally, not all combinations of
$\alpha_1$ and $\alpha_2$ are admissible. In general, there exists a
boundary $\alpha_2\leq f(\alpha_1)$ with $f(\alpha_1)\approx\alpha_1$
for $\alpha_1\approx 1$. In sparse hypergraphs, we typically have
$M\sim N^{1+\delta}$ with $\delta$ small, often $\delta=0$. Locally
however, the edge density can be much higher. For instance in ordinary
edge clustering $m\sim n^2$ and in triangle-based clustering $m\sim
n^3$, for $n$ not too large. Hence as $\alpha_1$ decreases from $1$,
the boundary function $f(\alpha_1)$ will deviate more and more from
the diagonal $\alpha_2=\alpha_1$. In Fig. \ref{fig:phase-diagrams}, we
have sketched a typical shape of a boundary function (thick line). At
$p=1$ (Fig. \ref{fig:phase-diagrams}, top left), the contour lines of
$s_p$ are straight lines and $s_p$ will clearly be maximal at small
values of $(\alpha_1,\alpha_2)$. As $p$ increases, the contour lines
become increasingly more concave, pushing the value where $s_p$
attains its maximum towards $\alpha_1=1$. For the idealized boundary
function in Fig. \ref{fig:phase-diagrams}, the transition is in fact
discontinuous and jumps from being at $\alpha_1=0.1$ (origin of the
axes) to $\alpha_1=1$ around $p=1.95$ [bottom left]).

Since the transition is in general sharp as a function of $p$ and can
even be discontinuous, we will in practice only use the default
edge-to-node ratio score with $p=1$ to identify dense hypergraph
clusters, or use a large value of $p$ (typically $p\gtrsim 10$) to
identify connected hypergraph components.

\subsection{Algorithm efficiency}
\label{sec:algorithm-efficiency}

For an undirected hypergraph with $N$ nodes, $M$ hyperedges and
maximum edge size $k_{\max}$, the update steps in the power algorithm
are at most of the order $k_{\max} M$ [eq. \eqref{eq:13}] and $N$
[eq. \eqref{eq:14} and \eqref{eq:15}]. The number of steps needed to
reach convergence depends on the convergence parameter $\epsilon$
[eq. \eqref{eq:16}] and therefore possibly also on the hypergraph
size. In practice, a maximal number of iterations $I_{\max}$ is
defined and convergence manually inspected when $I_{\max}$ is
exceeded. Determining the optimal threshold value is at most of the
order $N$ (number of possible threshold values) times $M$ (calculation
of the edge-to-node ratio score). Taken together, runtime is bounded
by
\begin{align*}
  t_{\text{run}} \leq I_{\max}\left[ O(k_{\max}  M) + O(N)\right] + O(MN).
\end{align*}

For directed hypergraphs, determining the optimal threshold pair over
all possible combinations of entries of the dominant singular vector
pair $(x,y)$ is of the order $N^2M$, which is often prohibitive. In
such instances, taking
\begin{align*}
  X_{\max} &= \argmax_{c} \R_{p,q}(u_{X_c}, y)\\
  Y_{\max} &= \argmax_{c} \R_{p,q}(x,u_{Y_c}),
\end{align*}
where we used the same notation as in Section
\ref{sec:spectr-clust-bicl}, results in an approximation which is
again $O(MN)$.

\section{Applications}
\label{sec:applications}

\subsection{Local and global alignment of complex networks}
\label{sec:local-glob-alignm}

The core idea for applying hypergraph clustering to the analysis of
edge-colored graphs is to translate the relation between multiple
interaction types (edge colors) into higher-order hypergraph edges. We
illustrate this idea by showing that local and global alignment of
complex networks with a bipartite many-to-many mapping between their
vertex sets can be naturally viewed as a hypergraph clustering
problem.

\begin{figure*}
  \centering
  \includegraphics[width=\linewidth]{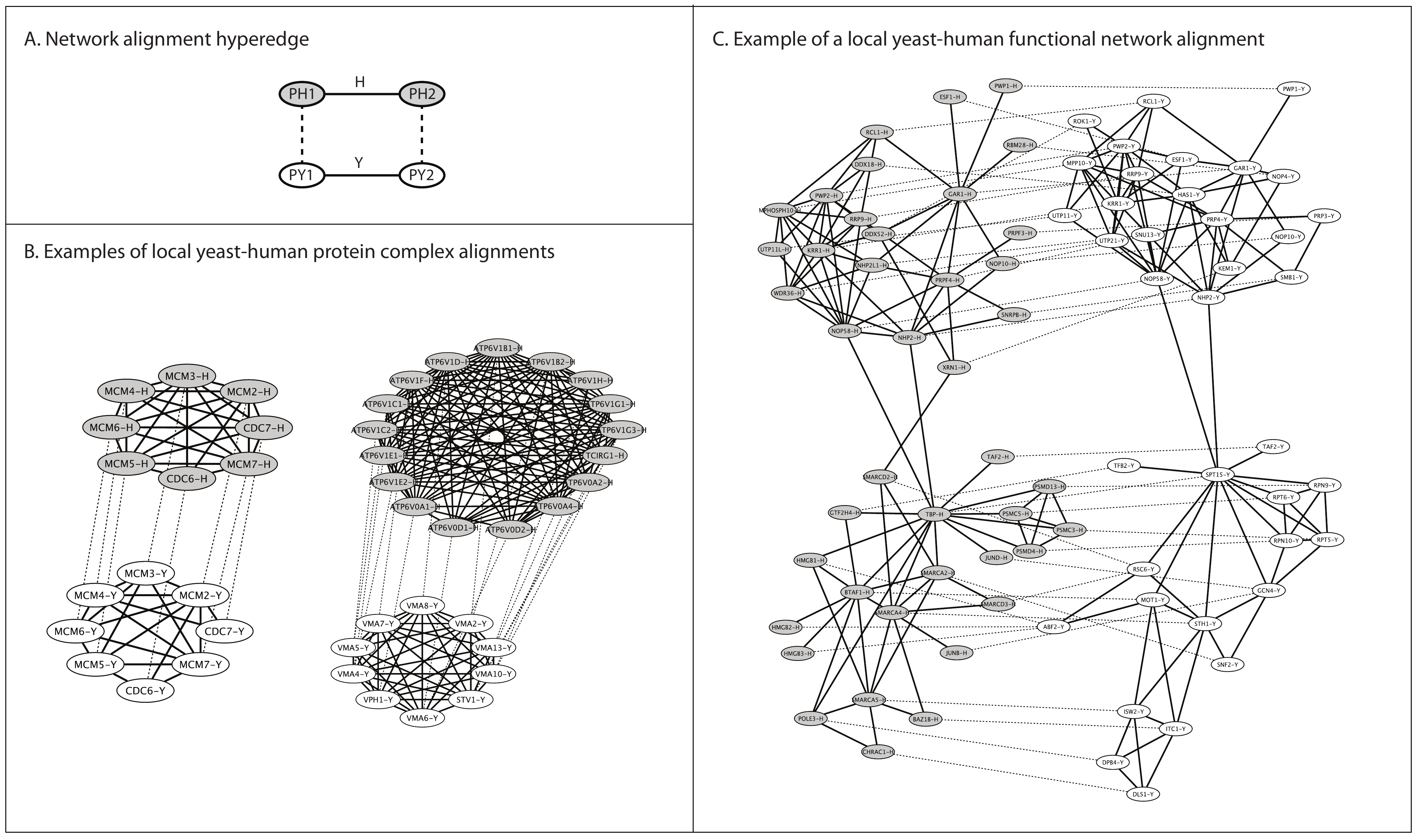}
  \caption{\textbf{A.} A (directed) hyperedge in the yeast-human
    protein interaction network alignment hypergraph is a so-called
    \emph{interolog}: a pair of interacting yeast (Y) proteins and a
    pair of interacting human (H) proteins connected by orthology
    relations (dashed lines). \textbf{B.} Examples of aligned protein
    complexes (cluster no. 19 left, no. 1 right). \textbf{C.} Example
    of a functional network alignment (cluster no. 48).  In all
    panels, yeast proteins are white and human proteins are grey;
    protein interactions are solid and orthology relations are
    dashed.}
  \label{fig:align}
\end{figure*}

Network alignment is the problem of finding topologically similar
regions between two or more networks. In local network alignment,
small subgraphs in each network are aligned independent of the
alignment of other subgraphs, whereas global network alignment aims to
find a maximal alignment for each connected component in the input
graphs. Network alignment methods for comparing molecular interaction
networks between different species come in two main
flavors. Topological network alignment finds conserved regions between
networks taking only the topology of each network into account
\cite{kuchaiev2010}. The second class of methods takes into account
that networks in different species have evolved from a common ancestor
through gene duplication and divergence mechanisms and hence that
there exists a meaningful mapping between the nodes in each network
\cite{sharan2006}. Methods have been developed which assume a
one-to-one mapping \cite{berg2006}, but in general a many-to-many map
should be considered \cite{sharan2004}. 

More formally, consider two ordinary graphs $\G_1$ and $\G_2$, whose
vertices are connected by a bipartite graph $\mathcal{M}$. The
directed alignment hypergraph $\mathcal{H}$ between $\G_1$ and $\G_2$
is defined as the $4$-uniform hypergraph containing the edges
$(\{i,j\},\{k,l\})$ if and only if $\{i,j\}\in\G_1$, $\{k,l\}\in\G_2$
and $\{i,k\}, \{j,l\} \in \mathcal{M}$ (Fig. \ref{fig:align}A).  Such
alignment hyperedges are also called \emph{interologs}. Interolog
mapping is routinely used to transfer annotation information from one
organism to another \cite{yu2004annotation} and interolog analysis is
at the heart of previous network alignment methods
\cite{kelley2003conserved,sharan2004}. Here we propose to address the
network alignment problem by identifying hyperedge clusters in the
alignment or interolog hypergraph. Indeed, in a local alignment, we
search for small regions in each graph which map nearly perfectly onto
each other, \textit{i.e.}, have a high density of interologs between
them. This corresponds to hypergraph clusters which maximize $\S_p$
for values of $p$ close to one. In a global alignment we search for
maximally matching regions in each graph, \textit{i.e.}, connected
components in the interolog hypergraph. These correspond to hypergraph
clusters which maximize $\S_p$ for large values of $p$.

We used our spectral clustering algorithm to locally and globally
align protein-protein interaction networks between yeast and human,
using orthology groups for mapping conserved proteins between both
organisms (see Appendix \ref{sec:alignm-yeast-human} for
details). Protein-protein interaction networks represent binary,
undirected associations between proteins and they are, at present, the
most extensively characterized molecular interaction networks in
biology \cite{barabasi2004,zhu2007}. Typical examples of high-scoring
local alignment clusters are conserved protein complexes \cite[Table
S2]{hypergraphclust_pre_supp}.  Fig. \ref{fig:align}B shows two
examples: first a set of proteins which map one-to-one between yeast
and human from the MCM complex (cluster no. 19), which plays an
important role in DNA replication and is indeed conserved among all
eukaryotes \cite{tye1999}; the second example (cluster no. 1) is a set
of components of the V-type ATPase (a proton pump) which has expanded
in human compared to yeast by gene duplications
\cite{kibak1992}. Other local alignment clusters reflect more general
functional networks than protein complexes \cite[Table
S3]{hypergraphclust_pre_supp}. Fig. \ref{fig:align}C shows cluster
no. 48, an example of a conserved network involved in nucleic acid
metabolism centered around the general transcription factor TBP (SPT15
in yeast), the TATA-binding protein. The largest connected component
in the network alignment hypergraph maps 651 yeast proteins to 766
human proteins and contains $90\%$ of all interologs \cite[Table
S4]{hypergraphclust_pre_supp}, showing that there exists a high degree
of network conservation at a global scale, consistent with previous
findings using topological network alignment \cite{kuchaiev2010}.

\subsection{Tripartite community detection in online folksonomies}
\label{sec:tripartite}

\begin{figure*}
  \centering
  \includegraphics[width=0.85\linewidth]{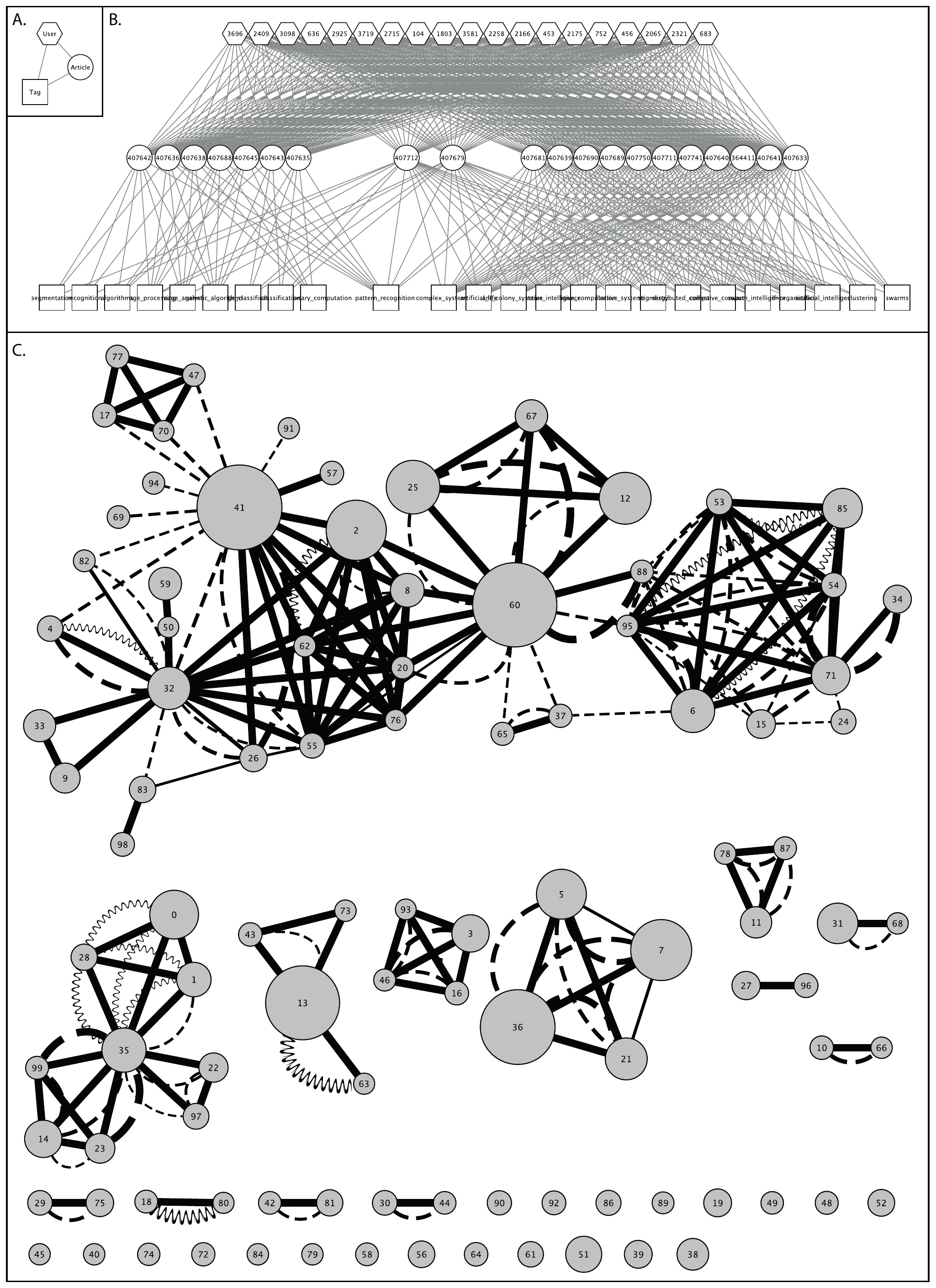}
  \caption{\textbf{A.} CiteULike hyperedge which represents one
    instance of a user (hexagonal node) who has annotated an article
    (circular node) with a certain tag (rectangular node). \textbf{B.}
    Example of two tripartite communities where the same set of users
    (top) has annotated two sets of articles (middle) with two sets of
    tags (bottom). Only the two central articles and one central tag
    (`pattern recognition') overlap between the two clusters. User-tag
    edges have been omitted for clarity. \textbf{C.} Coarse-grained
    view of the CiteULike hypergraph using the 100 highest-scoring
    hyperedge clusters. Each node represents a cluster (node size
    proportional to number of hyperedges in the cluster) and edges
    represent significant overlap between clusters (overlap score
    $>0.5$, edge size proportional to overlap score). Solid edges,
    user overlap; dashed edges, tag overlap; wavy edges, article
    overlap.}
  \label{fig:citulike}
\end{figure*}

Folksonomies, online communities where users collaboratively create
and annotate data, are examples of social systems that cannot be
adequately modeled by ordinary graphs. For instance, tagged social
networks such as Flickr \cite{flickr} or CiteULike \cite{citeulike}
have a tripartite structure that is best modeled by a 3-uniform
hypergraph \cite{ghoshal2009random, zlatić2009hypergraph}. Using
CiteULike as a concrete example, each hyperedge consists of a user who
has annotated an academic article with a certain keyword or tag
\cite{citeulike} (Fig. \ref{fig:citulike}A). Traditionally, the
community structure of such tripartite networks has been analysed by
considering one-mode ordinary graph projections of the hypergraph,
\textit{e.g.} by connecting two users if they have annotated the same
articles or connecting two tags if they have been applied to the same
articles \cite{zlatić2009hypergraph}. In contrast, hypergraph-based
clustering preserves the tripartite structure of folksonomy data and
reveals additional levels of community structure. We applied our
spectral clustering algorithm to a subset of the CiteULike data set
containing more than 400,000 (user, article, tag) entries and
identified nearly 14,000 hyperedge clusters (see Appendix
\ref{sec:trip-comm-detect} for details). The additional level of
detail present in hyperedge clusters is illustrated by looking at the
user, article or tag overlap between
clusters. Fig. \ref{fig:citulike}B shows an example of two hyperedge
clusters formed by the same set of users who have annotated different
sets of articles by different sets of tags. Only one tag, `pattern
recognition', is common between both clusters. The remaining tags show
that the articles in the first cluster are about collective computing
and swarm intelligence, whereas those in the second cluster deal with
image analysis \cite[Table S1]{hypergraphclust_pre_supp}, which are
indeed two distinct subjects within the broad field of pattern
recognition.

In general, we expect such sub-divisions of one-mode projected
communities to occur at the level of users (\textit{i.e.} the same set
of users annotating different sets of articles using different sets of
tags), but much less at the level of articles or tags (\textit{i.e.}
we do not expect different sets of users to annotate the same set of
articles using different sets of tags, or to use the same set of tags
for different sets of articles). Indeed, the 100 highest-scoring
clusters (which together contain about 20\% of all hyperedges) overlap
predominantly at the user level, to a much lesser extent at the tag
level, and hardly at the article level, while about 21 of these
clusters do not have any significant overlap (overlap $> 50\%$, see
Appendix \ref{sec:trip-comm-detect} for details) with any other
cluster (Fig. \ref{fig:citulike}C). Significant article overlap occurs
in only two instances. In both cases, it concerns a subset of users
who have annotated a subset of articles from a larger cluster with an
additional set of tags. Tag overlap occurs more frequently than
article overlap, but with lower overlap percentages than user
overlaps. Overlapping tags are typically general tags which can be
applied to a broad spectrum of articles. For instance, the ten tags
occurring most frequently in the top 100 clusters are: bibtex-import,
learning, social, evolution, review, support, govt, non-us,
collaboration, design. Thus we conclude that hyperedge clusters
capture topic-specific tripartite (user, article, tag) communities
which reveal more structure of the underlying data than user, article
or tag communities based on a single data-dimension only.

\subsection{Path clustering in regulatory networks}
\label{sec:path-clust-regul}

Unlike protein-protein interaction networks, which are undirected,
regulatory networks, which control the cellular response to external
or internal perturbations, are directed and represent the flow of
information within a cell \cite{alon2007b}. In transcriptional
regulatory networks, the response to perturbations can be measured
experimentally by genetically knocking-out a transcription factor (TF)
and measuring the resulting changes in gene expression levels on a
genome-wide scale \cite{hu2007}. In yeast, direct physical binding
interactions between a TF and its target genes \cite{harbison2004} as
well as perturbational response data for the same TF \cite{hu2007} are
available for a comprehensive set of almost 200 TFs (see Appendix
\ref{sec:path-clust-yeast} for details). On average only $~3\%$ of the
genes which respond to a knock-out perturbation of a TF are also
direct physical targets of that TF and various approaches have been
proposed to understand the mechanisms of indirect regulation and
propagation of network perturbations in this context
\cite{ideker2002,workman2006systems,gitter2009backup,joshi2010}. It is
thought that perturbational responses are organized in a modular way,
in the sense that groups of genes will be affected by the knock-out of
a TF through the same intermediate regulatory pathways. However, due
to the variable length of these pathways, previous approaches for
clustering in directed networks (\textit{e.g.},
\cite{guimera2007,palla2007,leicht2008}), which identify densely
interacting node sets, are not directly applicable to this problem.

\begin{figure*}
  \centering
  \includegraphics[width=0.8\linewidth]{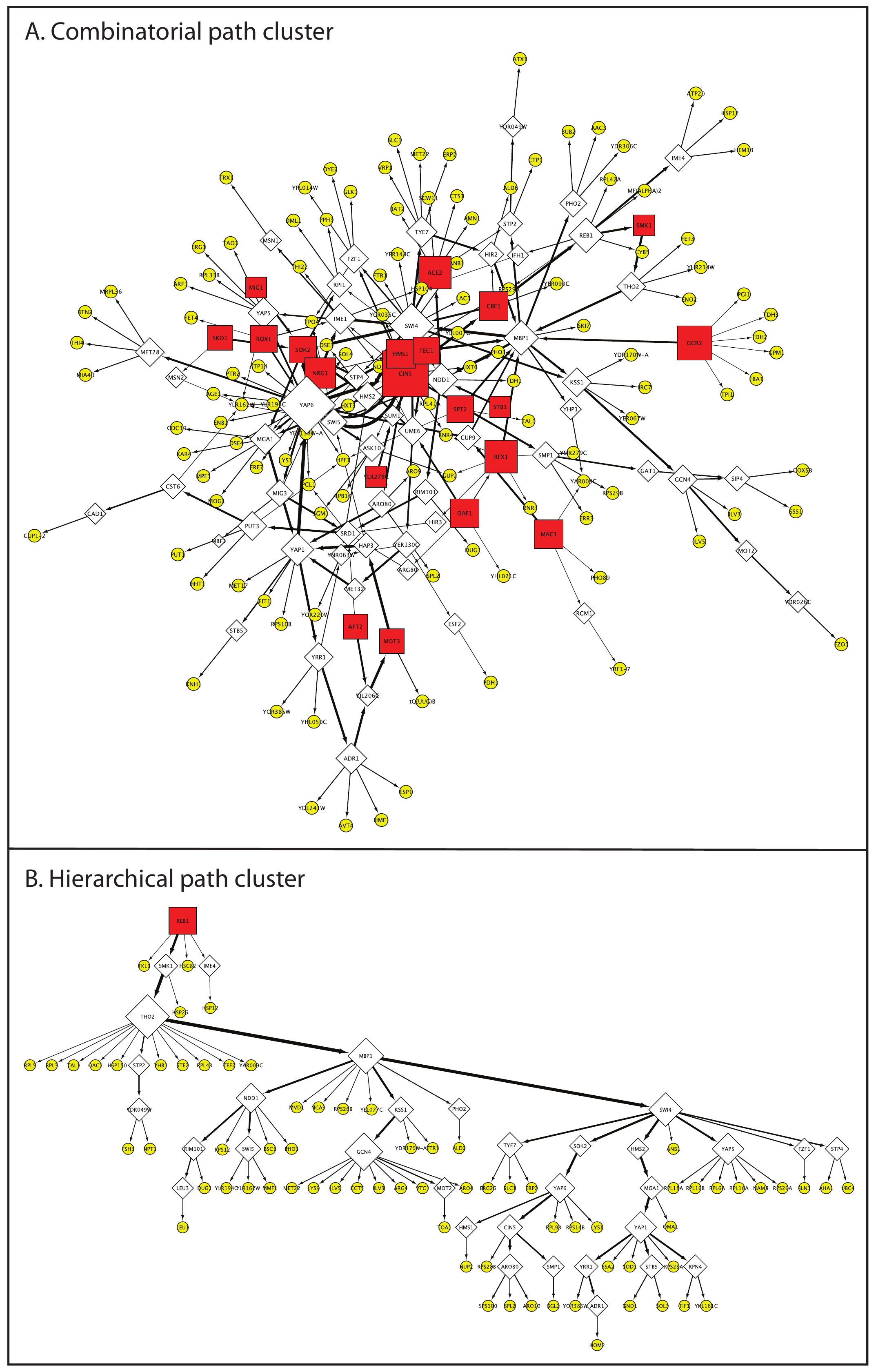}
  \caption{(Color online) Examples of a high-scoring combinatorial
    (\textbf{A}, Cluster no. 6) and hierarchical (\textbf{B}, Cluster
    no. 1) path clusters in the yeast transcriptional regulatory
    network. Red (dark gray) rectangular nodes, knocked-out
    transcription factors (TFs); yellow (light gray) circular nodes,
    genes differentially expressed upon knock-out of the TFs; white
    diamond-shaped nodes, all other TFs. Node size, resp. edge width,
    is proportional to out-degree, resp. edge betweenness (defined for
    the purposes of this figure as the number of shortest paths
    between all pairs of cluster nodes passing through a given edge).}
  \label{fig:paths}
\end{figure*}

Here we address the problem of identifying sets of nodes which respond
to the knock-out of a TF through similar regulatory paths by defining
a non-uniform hypergraph where each hyperedge corresponds to a
shortest path between two nodes in the original regulatory
network. Hypergraph-based clustering will then find sets of nodes with
a high number of shortest paths running through them and such clusters
form potential `signal-propagation' modules, consistent with the
notion that high information flow in a network is associated to high
values of a node's betweenness centrality (defined as the number of
shortest paths between all pairs of nodes passing through a given
node).  To test this hypothesis, we calculated all directed shortest
paths in the regulatory network of yeast between a TF and the genes
differentially expressed upon knock-out of that TF. The resulting
hypergraph contained 1332 hyperedges between 788 nodes and spectral
clustering identified 25 non-singleton and 14 singleton clusters (see
Appendix \ref{sec:path-clust-yeast} for details).  Topologically,
there appear to exist two distinct types of path
clusters. Combinatorial path clusters contain genes responding to the
knock-out of multiple TFs and form a network of densely overlapping
paths. Fig. \ref{fig:paths}A shows a combinatorial cluster of 199
shortest paths from 20 TFs to 186 genes involved in glycolysis and
gluconeogenesis. Hierarchical path clusters have a layered structure,
where the perturbational signal of usually not more than one TF flows
to its targets via a limited number of intermediate TFs, in a strictly
hierarchical manner (Fig. \ref{fig:paths}B). The functional relevance
of regulatory path clusters is demonstrated by the fact that they
contain a significant fraction of the genes affected by the deletion
of the cluster's TF and that they strongly overlap with specific
functional categories \cite[Table S5 and
S6]{hypergraphclust_pre_supp}.  For simplicity, we considered here
only shortest paths in the transcriptional regulatory network, but
clearly the approach can be extended to paths composed of multiple
interaction types.

\section{Conclusions}
\label{sec:conclusions}

Over the past decade, graph theory has become crucial to represent and
reason about complex network data. In particular clustering, the
detection of densely interconnected groups of vertices with few
connections to the rest of the network, has become a standard
coarse-graining procedure to understand the structure and function of
complex networks. With more and more data becoming available to
highlight different aspects of the same complex systems, a need has
arisen to analyze networks with multiple types of interactions
simultaneously. In this paper, we have proposed to use hypergraphs to
characterize higher-order relations between simple graphs and we have
introduced efficient algorithms for clustering and biclustering in
such hypergraphs.

Our main result is a spectral clustering algorithm for hypergraphs,
based on a generalization of the Perron-Frobenius theorem for directed
and undirected hypergraphs. More precisely, we have shown that, like
in ordinary graphs, there exists a unique, positive vector, called the
dominant eigenvector, over the set of vertices of a hypergraph, which
maximizes a natural generalization of the Rayleigh-Ritz ratio for
matrices. The importance of this result lies in the fact that the
ratio of the number of edges to the number of nodes in any subset of
vertices can be expressed as the same Rayleigh-Ritz ratio, in graphs
and hypergraphs alike. Densely interconnected clusters can therefore
be found very efficiently by first computing the dominant eigenvector
and then converting it to a discrete set of vertices. Uniqueness of
the dominant eigenvector guarantees unambiguity of the solution and
rapid convergence of the numerical procedure, whereas positivity
implies that the discretization can be achieved by setting an optimal
threshold on its entries.

Our work has been motivated by concrete problems of data integration
in social and biological networks. We have given three practical
examples for using hypergraph-based clustering in these contexts,
namely the alignment of protein-protein interaction networks between
multiple species using interolog clustering, the detection of
tripartite communities in folksonomies and the identification of
overlapping regulatory pathways in perturbational expression data
using shortest path clustering. Undoubtedly, many more applications
for hypergraph-based clustering exist in the analysis of other
biological, social, computer, communication or neural networks. From a
theoretical point of view, we have considered the edge-to-node ratio
as a simple quality score for clusters in graphs and
hypergraphs. Although this score has many attractive properties, such
as its direct relation with the dominant eigenvector and the absence
of any resolution limit problems, it will still be of interest to
generalize clustering algorithms based on other quality scores from
graphs to hypergraphs as well. Popular methods like those based on
minimal cutsets or modularity maximization also rely on spectral
properties of, respectively, the graph Laplacian and modularity
matrix. Although certain mathematical aspects, such as eigenvalue
multiplicity and its implications on algorithm convergence and cluster
discretization, are more complicated in these cases, we believe our
work lays the theoretical foundations for future studies in this
direction.

\appendix

\section{Strong connectivity of directed hypergraphs}
\label{sec:strong-conn-direct}

Consider first an undirected hypergraph $\Hg=(\V,\E)$ on $N$
vertices. Although connectedness of $\Hg$ does not imply
irreducibility, we do have the property that if there exists a proper
subset $I\subset \V$ such that for all $i_1,\dots,i_k\in I$ and
$j_1,\dots,j_m\not\in I$,
$w\bigl(\{i_1,\dots,i_k,j_1,\dots,j_m\}\bigr)=0$, then $\Hg$ is not
connected (since there can then be no path that starts in $I$ and
escapes from $I$). Hence if $\Hg$ is connected, no such set $I$
exists.

For a directed hypergraph $\Hg=(\V,\E)$ we can define an underlying
undirected hypergraph $\tilde \Hg=(\V,\tilde\E)$ by considering all
possible partitions of a subset $E\subset \V$ into source and target
sets, \textit{i.e.} $\tilde w(E) = \sum_{\{(S,T)\colon S\cup T=E\}}
w(S,T)$. This procedure generalizes the definition of a symmetric
adjacency matrix $B=A+A^T$ from the asymmetric adjacency matrix $A$ of
a directed graph. Clearly, to call $\Hg$ connected, we shall ask that
$\tilde \Hg$ is connected as defined in Section
\ref{sec:graphs-hypergr-tens}.

Now consider two subsets $I,J\subset\V$ such that $I\cup J$ is neither
empty nor equal to $\V$. Since $\tilde \Hg$ is connected, there exists
vertices $i_1,\dots,i_k\in I$, $j_1,\dots,j_\ell\in J$ and
$h_1,\dots,h_m\not\in I\cup J$ such that
\begin{align*}
  \tilde w\bigl(\{i_1,\dots,i_k,j_1,\dots,j_\ell,h_1,\dots,h_m\}\bigr)>0.
\end{align*}
This implies that there exists at least one partition of these nodes
into a source and target set with non-zero directed weight. We ask
slightly more, namely that there is a partition of the form
\begin{align*}
  w\bigl(\{i_1,\dots,i_k,h_1,\dots,h_n\},\{j_1,\dots,j_\ell,h_{n+1},\dots,
  h_m\}\bigr) > 0,
\end{align*}
\textit{i.e.}, the source as well as the target set should contain at
least one element not in $I$ or $J$. Note that the requirement that
all $i$'s go into the source set and all $j$'s into the target set is
purely notational convenience, since $I$ or $J$ are allowed to be
empty, as long as their union is not. If the above condition is
fulfilled for all pairs of sets $(I,J)$, we say that the directed
hypergraph $\Hg$ is \emph{strongly connected}.

\section{General proof of the Perron-Frobenius theorem for
  connected hypergraphs}
\label{sec:general-proof-perron}

Consider a non-negative maximizer $x$ of $\R_p(x)$ and without loss of
generality assume $\|x\|_p=1$. Let again $I=\{i\in\V\colon x_i=0\}$
and assume $I\not=\emptyset$. Let $k$ be the smallest integer for
which there exists at least one set $i_1,\dots,i_k\in I$ and at least
one set $j_1,\dots,j_m\not\in I$ such that
$w\bigl(\{i_1,\dots,i_{k},j_1,\dots,j_m\}\bigr)>0$. Such $k$ must
exists, since $\Hg$ is connected (see Appendix
\ref{sec:strong-conn-direct}). For $\epsilon>0$, define
\begin{align*}
  \tilde x_i =
  \begin{cases}
    x_i & i\not\in I\\
    \epsilon & i\in I
  \end{cases}
\end{align*}
We will show that for $\epsilon$ small enough, $\R_p(\tilde x)>
\R_p(x)$, contradicting the assumption that there can exist a
maximizer with zero elements. We have
\begin{align*}
  \|\tilde x\|_p^p = \|x\|_p^p + |I|\epsilon^p = 1 + |I|\epsilon^p,
\end{align*}
or, to leading order in $\epsilon$,
\begin{equation}\label{eq:3}
  \frac1{\|\tilde x\|_p} = 1 - \frac{|I|}{p}\epsilon^p + o(\epsilon^p).
\end{equation}
For the denominator of $\R_p(\tilde x)$, we have
\begin{align}
  &  \sum_{E\in\E} w(E) \Bigl( \prod_{i\in E} \tilde x_i
  \Bigr)^{\frac1{|E|}} \nonumber \\
  &\quad= \sum_{\{E\in\E\colon E\cap I=\emptyset\}} w(E) \Bigl(
  \prod_{i\in
    E} x_i \Bigr)^{\frac1{|E|}} \nonumber \\
  & \quad \qquad+ \sum_{\{E\in\E\colon E\cap I\not=\emptyset\}} w(E)
  \Bigl( \prod_{i\in
    E} \tilde x_i \Bigr)^{\frac1{|E|}} \nonumber \\
  &\quad= \R_p(x) + \sum_{\{E\in\E\colon E\cap I\not=\emptyset\}} w(E)
  \Bigl( \prod_{i\in E} \tilde x_i \Bigr)^{\frac1{|E|}}.\label{eq:4}
\end{align}
From the preceding discussion, it follows that the leading term in
$\epsilon$ of the second term in eq. \eqref{eq:4} is of the order
$\epsilon^{\frac{k}{k+m}}$ for some $k,m\geq 1$. Hence, for $\epsilon$
small enough, the extra positive term of order
$\epsilon^{\frac{k}{k+m}}$ in eq. \eqref{eq:4} offsets the negative
term of order $\epsilon^p$ in eq. \eqref{eq:3}, and we get, for some
$c>0$,
\begin{align*}
  \R_p(\tilde x) &= (1 - \tfrac{|I|}p \epsilon^p + o(\epsilon^p)) \bigl( \R_p(x) + c
  \epsilon^{\frac{k}{k+m}} + o(\epsilon^{\frac{k}{k+m}}) \bigr) \\
  &= \R_p(x) + c \epsilon^{\frac{k}{k+m}} +
  o(\epsilon^{\frac{k}{k+m}}) \\
  &> \R_p(x).
\end{align*}

Having established that a maximizer $x$ must be positive, $x>0$, the
remainder of the proof is the same as the proof for irreducible
hypergraphs, since in eq. \eqref{eq:2}, it suffices that at least one
$j_m\not\in I$ to arrive at a contradiction, which is guaranteed by
the connectedness of $\Hg$. 

For directed hypergraphs, the condition (and definition) of strong
connectivity in Appendix \ref{sec:strong-conn-direct} is tailor made
to ensure that the above argument still goes through. More precisely
if $(x,y)$ are a pair of non-negative maximizers of $\R_{p,q}(x,y)$
(cf. eq. \eqref{eq:7}), define $I=\{i\in\V\colon x_i=0\}$ and
$J=\{j\in\V\colon y_j=0\}$. Setting the zero-elements in $x$ and $y$
to a small positive value $\epsilon$, strong connectivity implies that
the numerator of $\R_{p,q}$ increases by a term of order
$\epsilon^\alpha$ with $\alpha<1$, whereas the denominator (the norms
of $x$ and $y$) can only decrease $\R_{p,q}$ by a term of order
$\epsilon^{\frac{p+q}{2}}$ with $p,q\geq 1$. The uniqueness argument
again follows along the lines leading to eq. \eqref{eq:2}.  \qed

\section{Network data and numerical settings}

Here we summarize the data sources and parameter settings used in the
example applications (Section \ref{sec:algorithm-validation} and
\ref{sec:applications}).

\subsection{Random geometric graphs}
\label{sec:geom-rand-graphs}

A geometric graph with $N$ vertices and radius $r$ is defined by a set
$\V$ of points in a metric space and edges $\E=\{(u,v)\in\V\colon
0<\|u-v\|\leq r\}$. We generated random geometric graphs by sampling
with uniform probability $N$ points in the unit square
$[0,1]\times[0,1]$ and taking the standard 2-norm as the distance
measure. For a given vertex, the probability that it is connected to
any other vertex is $\pi r^2$. Hence if we increase $N$ while keeping
$\rho=Nr^2$ constant we obtain a sequence of random geometric graphs
with constant average expected degree.

\subsection{Alignment of yeast and human PPI networks}
\label{sec:alignm-yeast-human}

We obtained physical protein-protein interactions (PPI) for yeast from
the BioGRID \cite{stark2006} database and physical and functional PPIs
for human from the BioGRID and STRING \cite{jensen2009string}
databases. The yeast network had 36,391 interactions between 4,847
proteins; the human network 40,630 interactions between 9,602
proteins.  We integrated these networks with orthology mappings from
the InParanoid database \cite{berglund2008}. There were 3,390
orthology relations between 2,245 yeast and 3,255 human proteins which
had at least one interaction in their respective PPI networks. We
performed recursive spectral clustering on the directed alignment
hypergraph consisting of 2,567 interolog-hyperedges
(cf. Fig. \ref{fig:align}A). At $p=q=1$, 180 clusters with at least
two hyperedges were found; 119 hyperedges had no connections in the
hypergraph, forming singleton clusters. The complete distribution of
hyperedges, nodes and scores for all clusters is shown in \cite[Fig
S2]{hypergraphclust_pre_supp}; the functional analysis of the local
and global alignment clusters is given in \cite[Table S2 and
S3]{hypergraphclust_pre_supp}.

\subsection{Tripartite community detection in the CiteULike data}
\label{sec:trip-comm-detect}

We obtained the complete `who-posted-what' data from CiteULike
\cite{citeulike} (\url{http://www.citeulike.org/faq/data.adp}),
containing (as of Feb. 1st, 2012) 16,553,642 (user, article, tag)
entries. To create a more manageable data set, we considered all
entries from 2005, resulting in a hypergraph of 466,948 (user,
article, tag) hyperedges between 4,693 users, 121,071 articles and
36,489 tags. Recursive hypergraph spectral clustering with $p=1$
identified 13,987 clusters with at least two hyperedges; 4,616
hyperedges formed singleton clusters. The complete distribution of
hyperedges, nodes and scores for all clusters is shown in \cite[Fig
S3]{hypergraphclust_pre_supp}. While comparing the user, article and
tag overlap of two hyperedge clusters, we were primarily interested to
detect when the set of users, articles or tags of a smaller cluster is
entirely contained in a larger cluster
(cf. Fig. \ref{fig:citulike}). We therefore used the overlap score
defined for two sets $X$ and $Y$ as
\begin{align*}
  \mathrm{ovlp}(X,Y) = \frac{\bigl|X\cap
    Y\bigr|}{\min\bigl(|X|,|Y|\bigr)},
\end{align*}
which reaches its maximum value of 1 whenever $X\subset Y$ or
$Y\subset X$.

\subsection{Path clustering in the yeast transcriptional regulatory
  network}
\label{sec:path-clust-yeast}

We obtained a network of 11,373 physical transcription factor (TF)
binding interactions between 198 TFs and 3,535 target genes in yeast
from \cite{harbison2004} and knock-out microarray data for 266 TFs
from \cite{hu2007}. The knock-out data can be represented as a
directed network of perturbational interactions where each TF is
connected to the genes which respond to the knock-out perturbation of
that TF. 182 TFs with physical binding data also had knock-out data
for a total of 7,090 perturbational interactions.  We constructed a
directed hypergraph consisting of 1,332 hyperedges and 788 nodes,
where each hyperedge is a shortest path in the regulatory network
between a TF and a gene differentially expressed upon knock-out of
that TF. We defined the source set of a hyperedge as the knocked-out
TF and the target set as the remainder of the path. Recursive spectral
clustering identified 39 clusters of which 14 were singletons.

\subsection{Supplementary data and algorithm implementation}
\label{sec:algor-impl}

An implementation of the clustering algorithm in Java, together with
the input data and clustering results described in Section
\ref{sec:applications}, is available from the project homepage at
\url{http://schype.googlecode.com}.

\begin{acknowledgments} TM thanks the Department of Mathematics of the
  University of California, Davis for warm hospitality during visits
  when part of this work was performed. The work of BN was supported
  in part by the National Science Foundation under grant
  DMS-1009502. We thank the referees for their detailed and
  constructive comments which lead to a much improved version of our
  paper.
\end{acknowledgments}


%

\end{document}